# Molecular Dynamics Simulation of Nanoporous Tungsten

Author names and affiliations:

Jarod Worden[*], Celine Hin[*,**]

*. ME department
Northern Virginia Center, Virginia Polytechnic Institute
7054 Haycock Rd, Falls Church, VA 22043, United States
**. MSE department
Northern Virginia Center, Virginia Polytechnic Institute
7054 Haycock Rd, Falls Church, VA 22043, United States

Corresponding Author
E-mail address: jaworden@vt.edu (Jarod Worden)
Postal Address: Northern Virginia Center, Virginia Polytechnic Institute
7054 Haycock Rd, Falls Church, VA 22043, United States

Abstract

   Developing materials that can withstand the intense environments of nuclear fusion reactors is critical in developing long-term commercial viability for energy production. Tungsten is the primary candidate as a plasma facing material due to its exceptional resilience in extreme fusion settings. However, the continuous high energy helium and neutron damage threatens mechanical stability, calling into question the long time commercial reliability for fusion energy generation. In order to address these issues, novel methods of nanoporous structures have been shown to suppress and annihilate the accumulation of defects that leads to structural instability that can be applied to improve the reliability of tungsten as a plasma facing material. There is currently a lack of experimental data on nanoporous Tungsten that needs to be considered when choosing appropriate applications for this material. Molecular dynamic simulations were utilized in examining the mechanical and thermodynamic response of nanoporous tungsten systems under fusion reactor relevant conditions under compressive strain. Bicontinuous nanoporous structured Tungsten with small relative densities are shown to have outstanding thermodynamic and mechanical stability under compression compared to other Tungsten systems by possessing reduced defect densities and unique defect twinning.

# I. Introduction

As the energy needs of the world grow, nuclear power will play an important role in supplying reliable, carbon-free energy. Newer designs of nuclear reactors will be needed to meet the demands of higher energy needs while reducing the environmental impact of nuclear waste. One of the most ambitious advances for meeting these requirements is the construction and demonstration of a stable nuclear fusion reactor. Theory and conceptualization for this type of reactor has brought much interest due to the potentially unlimited energy supply. Many different fusion test reactors, such as ITER in France and SPARC reactor from MIT, are nearing completion and the demonstration phase for prolonged nuclear fusion energy generation [1-4]. However, one of the main issues is selecting appropriate materials that can withstand the high-energy neutron and helium bombardment of the confined plasma in the reactor environment [5-7]. Tungsten (W) has been selected as one of the primary candidate materials for many applications in the fusion reactor due to its high melting point temperature, high strength, and high displacement energy that makes it resistant to radiation damage [8,9]. Although Tungsten has a high resistance to radiation damage, defects are still generated in the material, leading to complex defect formations that could lead to structural instability [9,10]. An increase in vacancies and self interstitial atoms (SIA) caused by neutron radiation damage are known to lead to hardening, embrittlement, and an increase in the brittle to ductile transition temperature [11,12]. Helium bombardment and accumulation in W have been shown to influence grain boundary cracking and surface coarsening [13,14]. Suppressing these defects is key to maintaining stable mechanical properties for the operation of the fusion reactor for long time periods. Material sinks, such as surfaces in nanoporous materials (NM), are known to control and suppress defect concentrations by removing the defects in the bulk [15,16]. NM have high surface to volume ratios that facilitate fast recombination of vacancies and SIAs compared to traditional bulk materials that should prolong the operational lifetime of the nuclear reactor [17-20]. The high surface area to volume ratio also removes He from the system more quickly than bulk or nanograined materials, reducing large bubble formations that could lead to bursting at the W surface and grains [21]. With these qualities of radiation resistance, Nanoporous Tungsten (NP W) could be material considered for nuclear fusion applications. In order to properly incorporate these materials into the operation of the nuclear environment, there needs to be an understanding of the mechanical properties of the material at the high operating temperatures. Due to the novelty of such an approach with NP W, there is a lack of experimental results for the mechanical properties. This paper's aim is to utilize Molecular Dynamics (MD) in predicting the mechanical response of NP W under uniaxial compression at nuclear fusion reactor temperatures and compare the results with polycrystalline and bulk W [22,23]. The types of NP W that will be investigated include nanopore insertion and bicontinuous nanoporous (b-NP) W. Tungsten with nanopores have been used to investigate irradiation resistance and b-NP W is another type of structure that is expected to have high irradiation resistance [20,24]. It is important to compare mechanical properties of these different types of nanoporous structures as they might produce different mechanisms that are preferred for fusion reactors.

# II. Simulation Details

MD simulations were investigated using the open source code LAMMPS [25]. The simulations were carried out in a three dimensional cell with periodic boundary conditions in the three directions ([100], [010], and [001]) with lengths of 39.56 nm x 39.56 nm x 39.56 nm. The bulk cell was produced by populating with 3906250 bcc W atoms. The open source code Atomsk is used to construct the nanocrystalline (NC) systems with 5.9/10/20 nm grain sizes by using Voronoi Tessellation on the bulk cell with random grain orientations [26]. The systems then undergo atom removal along grain boundaries where atoms within 2 Angstrom cutoff range are removed. The Polak-Ribiere version of the conjugate gradient algorithm is used to reach the minimum energy configuration for the grains in the NC system. NC W is visualized for 10 nm grain in Fig. 1a. Two separate methods of producing NP systems are considered. The first method involves implanting spherical pores in the system with different relative densities, pore sizes, and pore distributions. This involves 50% and 70% relative densities with either one large, central pore or 16 smaller equidistant pores which are shown in Fig.1b and Fig.1c. The second method is constructing a b-NP structure that was produced following similar methods used in modeling b-NP gold through dealloying simulation methods [27]. The demixing was conducted to produce 28.0/37.2/46.5/56.1/68.7% relative density b-NP W using the bcc W-Ta system that utilizes an EAM W-Ta potential with an imposed repulsive Lennard-Jones potential [28].

The demixing was performed through an NVT ensemble at temperatures above melting point (3700 K) for a set amount of time to allow the formation of separate phases. After the set amount of time, the Ta phase is deleted with only the W atoms remaining. The remaining W atoms are replaced with perfect crystal bcc W atoms such that the b-NP structure remains. The system is then relaxed in an NPT ensemble at room temperature and atmospheric pressure. Fig. 2 shows the simulation cell of 28.0% and 68.7% relative densities for b-NP W. The average ligament diameter was calculated with a homemade procedure where atom locations are selected to be the center of an expanding sphere. The sphere expands with a radial increase of half the lattice parameter and the surface detects the local atom configuration to determine if a section of the sphere is in contact with the surface of the ligament. When the ligament surface is detected, the opposite side of the sphere is checked to see if it is also in contact with a surface. If it is, the distance is calculated as the diameter for a section of the ligament that is averaged with the other diameters that were calculated. The b-NP system's average ligament size and relative density are shown in Table 1.

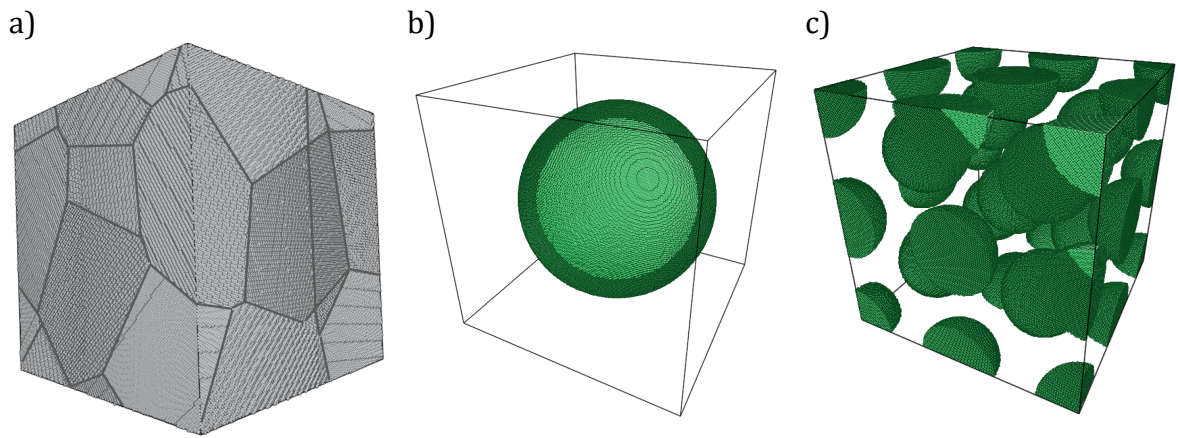

**Figure 1**. W systems that represent (a) 10nm grain W, (b) 70% relative density single large pore, (c) 70% relative density multiple pores.

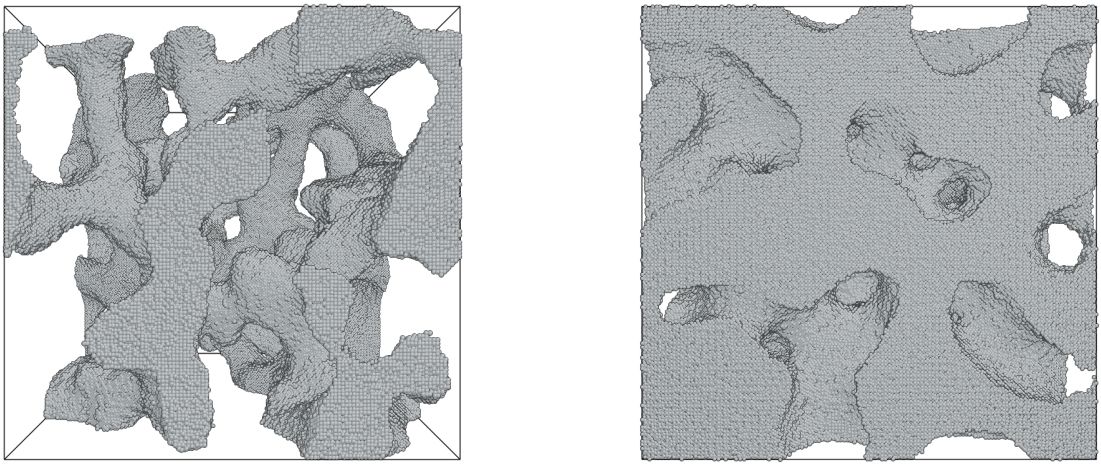

**Figure 2**. The 28.0% (left) and the 68.7% (right) relative density systems of b-NP W

**Table I**. The relative density with the average ligament size of the system.

| Relative Density | Ligament Size (nm) |
|:---:|:---:|
| 28.0% | 4.69 |
| 37.2% | 5.44 |
| 46.5% | 5.78 |
| 56.1% | 6.11 |
| 68.7% | 7.78 |

### III. Results and Discussion

This section is used to show the results for the simulation models under compression for bulk, NC, nanoporous, and b-NP W. The mechanical compression tests are performed at multiple temperatures from room temperatures to fusion reactor operating temperatures [29]. A Finnis-Sinclair potential is used to model the atomic interactions between the W atoms that is fitted to a mixed database of experimental and ab initio calculations [30]. The same NPT setup used for thermalization is performed on the system at the chosen temperature for 100 picoseconds to achieve energy minimization. A strain rate of $5.0 \times 10^9$ s$^{-1}$ is imposed to deform the simulation box in the [100] direction. The large strain rate is a limitation of the small time scales of MD simulations where realistic strain rates would require large computational cost and time. Defects in the system were analyzed using the BCC defect analysis tool (BDA) and the Dislocation Analysis Tool (DXA) [31,32]. Both the tools are used for analyzing the defect types and concentrations during the strain process. The BDA tool is more robust compared to the DXA tool and is used for analyzing the specific defects identified in bcc lattice structures such as vacancies, dislocations, twin boundaries, surface atoms, and planar faults. The DXA tool is used to assist in identifying specific Burgers vectors of the dislocations.

#### A. Stress-Strain Curve Analysis

The stress-strain curves of the W systems under 298/600/900/1150 K are shown in Fig. 3. For initial strain values between 0% and 2.5%, bulk and NC systems display an initial elastic region of linearly increasing stress until yield strength is reached. When strain continues to increase, NC systems undergo plastic deformation where the increase in stress changes from linear to logarithmic growth for strain values between 2.5% and 6% until ultimate stress is achieved. Stress then decreases as strain increases until fracture occurs and the stress remains constant for increasing strain values. The bulk responds differently by displaying a longer elastic region that continues until a high yield stress is reached at 17.5%. This compressive response is observed in other W molecular dynamics models that display high yield stress and yield strain values compared to experiment due to the high activation energy of dislocation formation that results in high strength and brittleness [33,34]. The bulk then undergoes failure with increasing strain, displaying a rapid decrease in stress. The nanoporous structures display different responses to compression compared to the bulk and NC systems. The nanoporous structures with spherical pores present two separate elastic regions with a plastic region in between them. The initial elastic region has strain values from 0% to 2.5% which leads to the intermediary plastic region with strain values 2.5% to 10%. The stress then increases back to an elastic region beginning around 11.5%. Larger pore systems display a higher ultimate stress and strain corresponding to ultimate stress compared to multiple pore systems. Higher ultimate stress is also found in systems with higher relative densities. Once this ultimate stress is achieved, the nanoporous systems undergo stress loss and fracture with increasing strain. The b-NP W system curves follow other b-NP systems for both MD simulations and experimental testing where there is a small elastic region at initial strain values, flattens for a large range of strain values, then exponentially increases at larger strain values representing strain hardening and densification [35–37]. This is compared with compression tests done on b-NP Au since it is the most heavily studied material for this purpose and there is a lack of b-NP metals with bcc crystal structure. The initial elastic region is slightly larger compared to the NC and nanoporous systems with strain values from 0% to 3.5% for the 68.7%. As relative density decreases, the elastic region increases with the 28.0% relative density system showing an elastic region of 7% strain.

The Young's modulus of the bulk W is 402.79 GPa at room temperature that compares well to the experimental value of 410 GPa [38]. The Young's modulus is calculated through linear regression up to 2% strain. The Young's modulus for all the systems at room temperature are calculated and provided in Table 2. The bulk W follows very closely with other MD simulations of W with similar ultimate stresses and strain fracture values [34]. When analyzing the NC W systems, the Young's modulus decreases with decreasing grain size that follows NC W simulations results [39]. The 5.9 nm grain size Young's modulus also matches closely with other MD simulations of 5.6 nm grain size Young's modulus calculations where Tahiri et al. obtained a Young's modulus of 360 GPa [40]. The b-NP W systems follow Gibson-Ashby Young's modulus scaling law for porous microstructures. The Young's modulus as a function of porosity is given as [41,42]:

$$Y^{eff} = Y^0 C_0 \varphi^m$$

In this equation, $Y^{eff}$ is the Young's modulus of the b-NP system, $C_0$ is a constant, $Y^0$ is the bulk Young's modulus, $\varphi$ is the relative density and $m$ is a constant. When fitting the Young's modulus to the scaling law, the parameters obtained are $C_0$ = 0.9 and m = 2.1 with a goodness of fit $R^2$ = 0.9942. When compared to nanoporous Au systems, the exponent agrees well with results obtained where the exponent is estimated to be 2 [43,44]. The prefactor is also close to results where the suggested value is 1 [41,44,45].

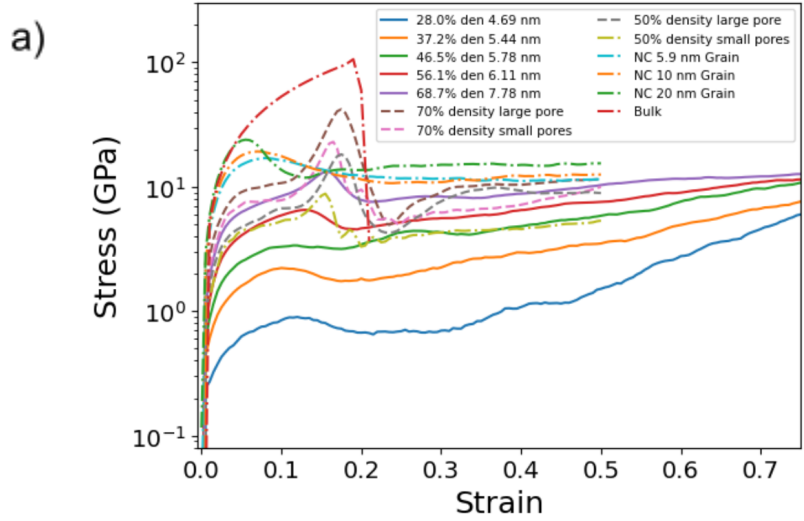
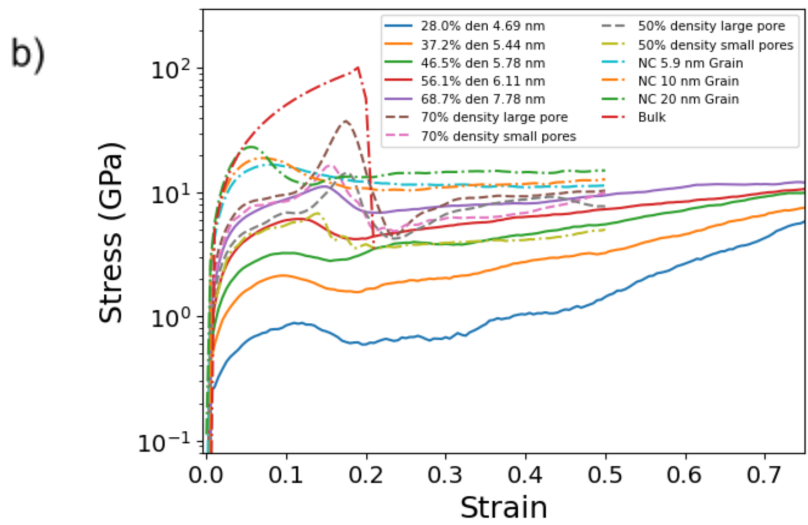

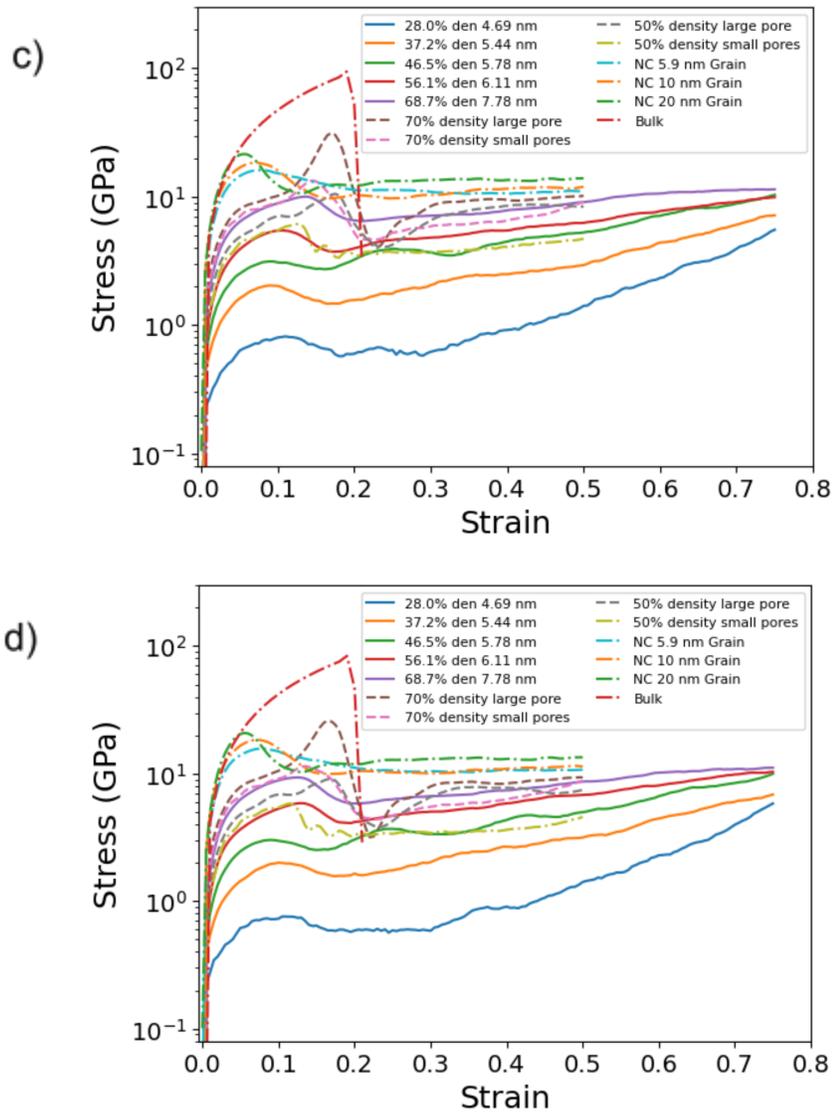

Figure 3: The stress-strain compression tests curves of the W systems for 298 (a) /600 (b) / 900 (c) /1150 (d) K.

**Table II.** The Young's modulus for given W systems at room temperature.

| System | Young's Modulus (GPa) |
|---|---|
| Bulk | 402.79 |
| NC W 5.9 nm grain | 368.68 |
| NC W 10 nm grain | 382.42 |
| NC W 20 nm grain | 396.11 |
| NP W 70% density large pore | 222.95 |
| NP W 70% density small pores | 184.72 |
| NP W 50% density large pore | 139.97 |
| NP W 50% density small pores | 112.37 |
| b-NP W 28.0% density | 18.85 |
| b-NP W 37.2% density | 46.75 |
| b-NP W 46.5% density | 71.83 |
| b-NP W 56.1% density | 110.45 |
| b-NP W 68.7% density | 173.64 |

Fig. 4 shows the negative change in ultimate stress as a function of temperature. In the b-NP W, the systems with lower relative density have a higher resistance to ultimate strength reduction when considering temperatures ranging from room temperature to higher operating temperature of 1150 K. The higher relative density systems also follow lower ultimate and yield stresses for higher temperatures. The higher thermal stability of the lower relative densities can be attributed to the smaller ligament sizes of the lower density system [46]. Higher thermal stability in lower relative density systems comes from the higher surface-to-volume ratio and stability in twin boundary concentrations that removes defects from the system, such as vacancies, much more quickly compared to the bulk matrix [24]. This can be seen when comparing 30% relative density and 70% relative density b-NP W at 298 K and 1150 K. Fig. 6 shows that there is minimal change in defect concentrations for the 30% relative density against varying temperature compared with 70% relative density. For the 70% density, there is an increase in planar fault and vacancy concentration for higher temperatures, lower twin boundary concentration, and no substantial change in dislocation density. Due to twin boundaries acting as sinks for vacancies to be taken out of the bulk, a lower twin boundary density in higher temperature regimes raises the vacancy concentration and increases the number of vacancy clusters which reduces vacancy mobility in the bulk to the surface [47,48]. The increase in vacancy concentration has a detrimental impact on the maximum stress as well as a decrease in the strain value that corresponds to the maximum stress [49]. A change in vacancy concentration appears to be a primary defect type that affects the stress strain curve when comparing the nanoporous systems against the bulk systems. Shen et al. found that the initial elastic region is insensitive to the number of vacancies until the maximum stress is reached which is observed by the larger amount of vacancies in the higher temperature system [49]. In bulk W, vacancies reduce the strength by offering more possibilities for bond breaking and dislocation to occur, however the dislocation density for 70% relative density systems is relatively the same through strain values under 35%, showing how dislocations are insensitive to vacancy concentration in b-NP W. This changes the primary mechanism for strength loss by vacancies in nanoporous materials as bond breaking. Another mechanism that appears to be important in thermal stability is the planar fault concentration increase in 70% relative density b-NP W at higher temperatures. Planar faults are observed in many different bcc materials under compression and can represent detrimental mechanisms in the system such as cracks and notches in the ligament structure that can weaken the structural stability [18,19].

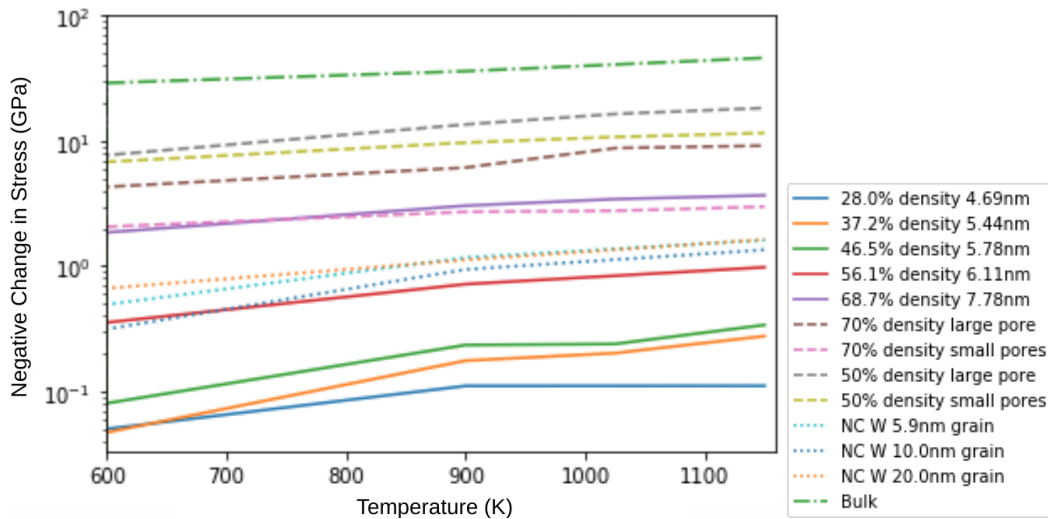

**Figure 4**. Graph showing negative change in ultimate stress with increasing temperature.

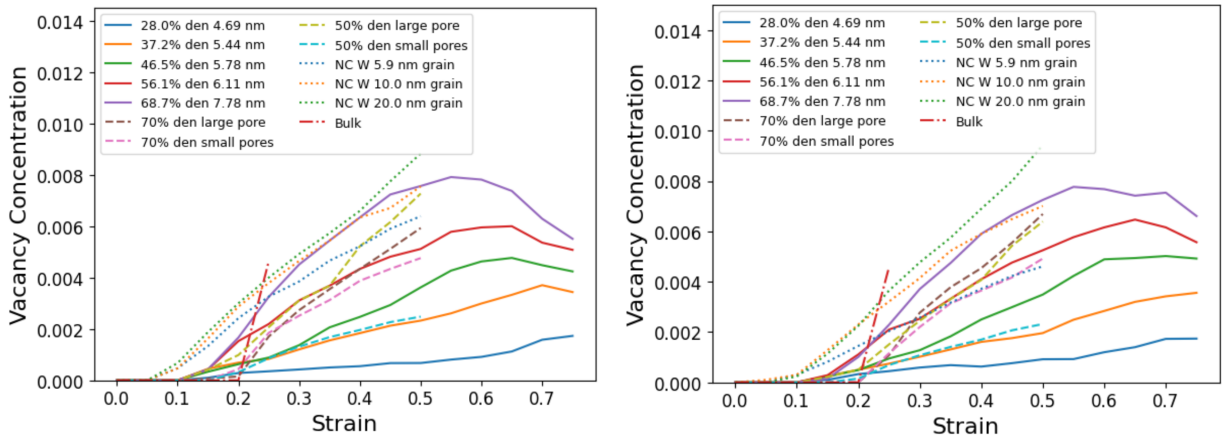

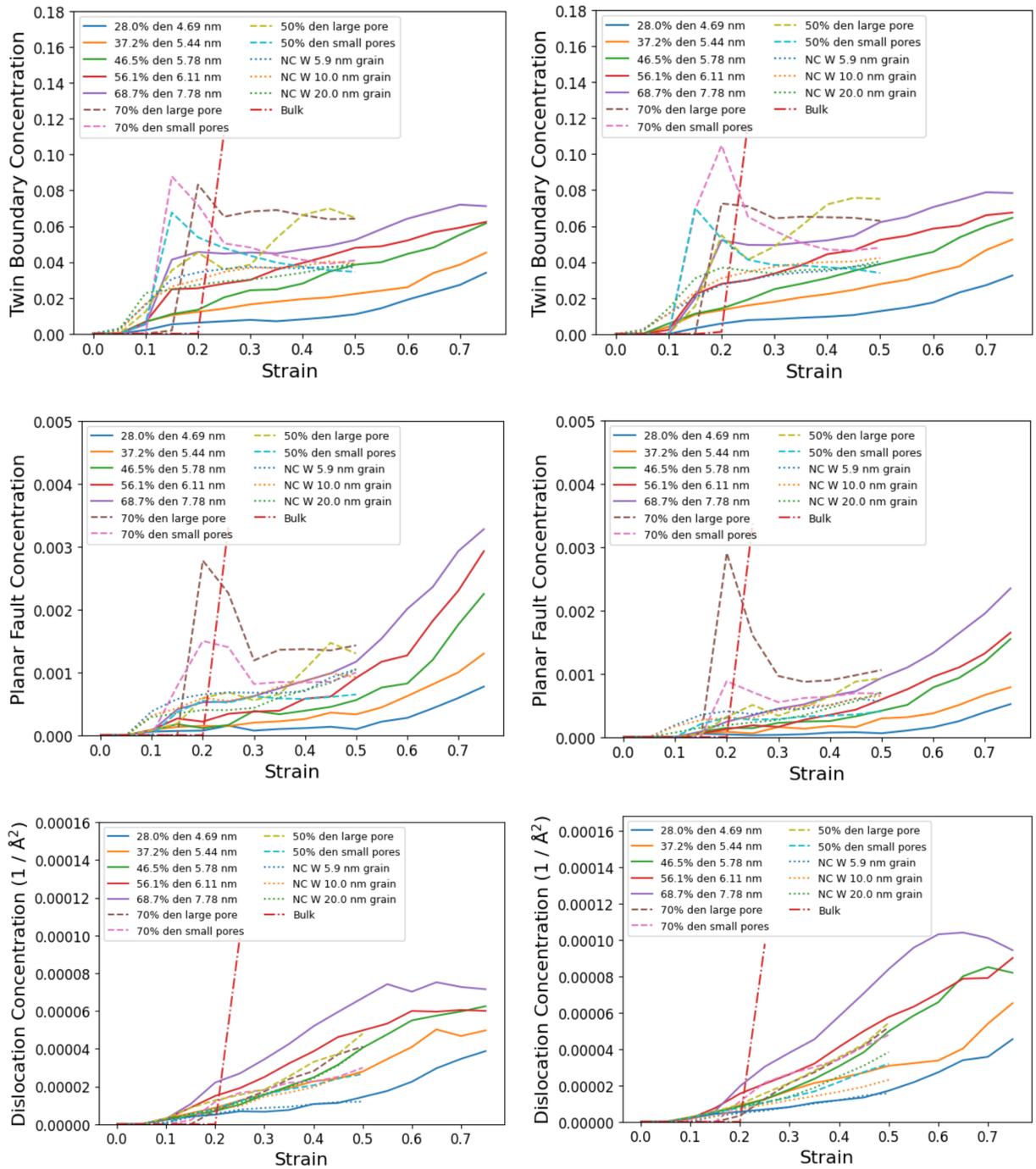

**Figure 5**. Defect concentrations of nanoporous, nanocrystalline, and bulk at 1150 K (left column) and 298 K (right column) for vacancies (first row), twin boundaries (second row), planar faults (third row), and dislocation (fourth row) densities.

## B. Defect Analysis

The BDA tool is used in identifying bcc defects including vacancy, vacancy clusters, twin boundaries, dislocations, and stacking faults. Ovito was used to display the defects obtained through the BDA tool where the dark blue represents surface atoms, the light blue represents dislocations, the yellow represents twin boundaries, the light red represents planar faults, and the dark red are unidentifiable defects which could correspond to a local phase change [50]. The DXA tool is used in order to more precisely identify dislocations in the structures compared to the BDA tool, which only identifies surrounding atoms making up a dislocation.

Fig. 5 shows the defect concentrations of the systems for room temperature and operating temperatures. When analyzing the dislocation densities, all the systems have reduced densities when comparing the 298 K to 1150 K. The reduced dislocation density can be attributed to the increased kinetics of dislocation motion in the material at higher temperatures through reduced internal friction, facilitating slip, and decreasing material flow stress that would increase dislocation annihilation [51,52]. The most common Burgers vectors identified by the DXA were in the 1/2<111> direction with the next most common being in the <100> and <110> directions. This matches well with analysis of W nanowires where 1/2<111> is a very common Burgers vector under compression [53]. The primary dislocation type is mixed dislocation as there is no clear distinction between the dislocation types in the b-NP and nanoporous systems. For the nanoporous systems however, the dislocations at the surface of the pores can be identified as dislocation loops in the 1/2<111> direction in good agreement with other simulations of porous bcc systems [54]. The vacancy concentrations for the systems increase as temperature increases due to thermal vacancies [55]. Concerning the lower density b-NP W systems, the vacancy concentration does not change nearly as much compared to other systems. This can be attributed to the higher sink density of the high surface-to-volume ratio that annihilates thermal vacancies from the bulk more quickly. Twin boundary concentration does not appear to be heavily impacted by temperature except for the 70% density pore system where the concentration decreases with increasing temperature. The change in planar fault concentrations is the most pronounced change in the systems with higher concentrations at increased temperatures.

Surface energy was calculated as a function of strain at operating temperatures and is displayed in Fig. 6. The difference in surface energy between the NP and b-NP systems may come from the different surface curvatures where the NP structure strictly contains concave curvature and the b-NP systems have a mix of concave, convex, and saddle curvatures. The b-NP systems begin with smaller surface energy compared to the NP systems and then increase with increasing strain. All the systems undergo sharp increases in surface energy from 0-20% strain values. The b-NP surface energy is dependent on the relative density where the higher relative densities have higher surface energies at all strain values. The higher relative densities also increase in surface energy the most compared to lower relative densities where the 28.0% relative density increases by 0.0159 eV/A$^2$ compared to an increase of 0.6086 eV/A$^2$ in the first 20% strain range for the 68.7% relative density. The NP W systems undergo some surface energy loss around 20%, but gradually increase surface energy as strain increases. When comparing the curves of the surface energy to the defect concentrations, the surface energy has similar trends with the planar fault concentration. This relation matches with faults in metals having surface energies of their own, increasing the overall surface energy of the system meaning that an increase or decrease of the surface energy would match an increase or decrease, respectively, of planar fault concentrations [56,57]. When looking at the curves for the surface energy, another correlation is that higher surface energies tend to lead to higher defect concentrations. Minimizing surface energy and maintaining low surface energy through the straining process will keep defect concentrations low which is a key aspect for thermal and mechanical stability [58].

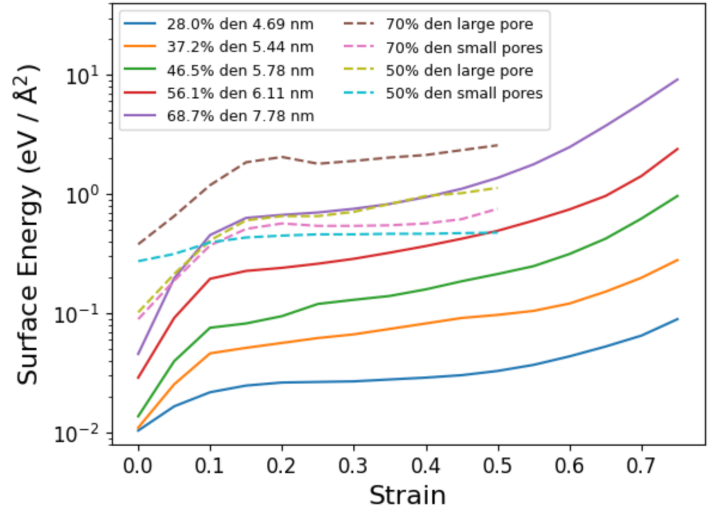

**Figure 6**. Change in surface energy as a function of strain at 1150 K. The solid lines represent b-NP W and dotted lines are the NP W.

Fig. 7 shows the BDA analysis for defects in a 70% relative density, single large nanopore under compression. The defects in the systems are represented by different color spheres where the surface atoms are dark blue, vacancies are light blue, dislocations are teal, twin boundaries are yellow, and planar faults are red. Perfect bcc atoms are not represented by any spheres. Lower strain values either do not produce any defect or low amounts of unidentifiable defects from 0-10%. The unidentifiable defects could be atoms that are not in bcc crystal structure. When analyzing the placement of the defects, there is a lack of defect accumulation in the volume that is parallel with the strain direction [100] and perpendicular with the pore surface in [010]. This persists in many increasing strain values. When the strain value gets to 15%, larger defects such as twin boundaries and dislocations begin to form around the pore and are restricted to the volume near the surface with some twin boundary formation occurring deep in the bulk. The twin boundary formation in the <111> direction and shear dislocations in the <111> direction begin to form on the points on the surface of the pore that correspond to the surface with slip planes in {211} and is shown in Fig. 8. The defect nucleation follows similar work in nanopore W and Ta systems that found the emission of shear loops are dependent on the schmid factor for specific slip systems [54,59]. The first signs of twin boundary and dislocation formation suggest the location where the pore surface deforms the ellipsoid shape inward by assuming that dislocation loops transport matter into the voids and is the first indication of void collapse [60]. The dislocations protrude further out from the pore surface into the bulk through help of twin boundaries on the surface. As strain increases, the dislocations and twin boundaries grow around the pore and into the bulk with the pore remaining as a nucleation site. This agrees well with models by Marian et al. that states dislocation loops extend into the bulk while remaining attached to the void surfaces under uniaxial compression [61]. At the strain value where the system begins to lose stress is demonstrated in Fig. 7c where the surrounding bulk collapses and with many dislocation and twin boundary formation. This also allows for the formation of planar faults and vacancies to form in the disordered volume. The volume that is parallel with the [100] strain direction and perpendicular to the pore surface in the [010] direction still remains in perfect bcc crystal structure through increasing strain until 40% strain is reached. At the 40% strain, dislocations can be seen in Fig. 7d to begin forming in the perfect bcc structure region connected to the disordered volume region.

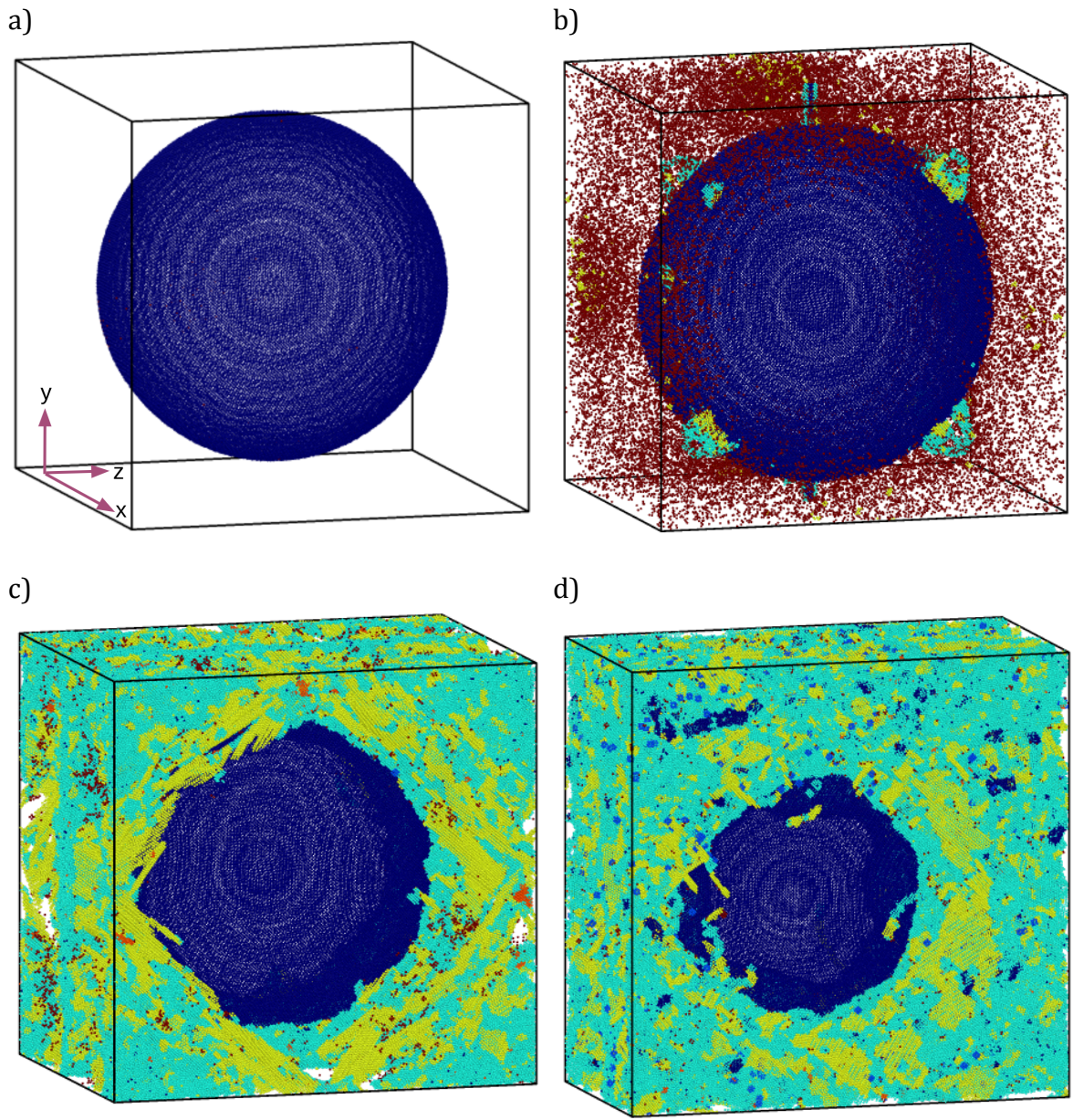

**Figure 7**. 70% relative density W with single large pore for strain values of (a) 0%, (b) 15%, (c) 20%, (d) 40%. Perfect bcc atoms are removed for visualization purposes.

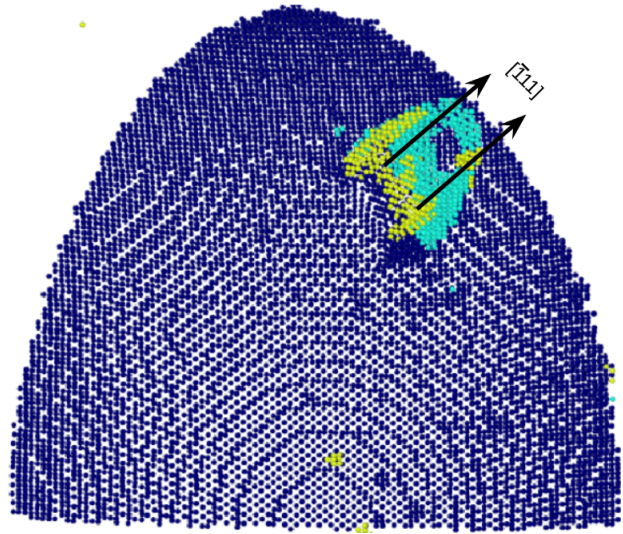

**Figure 8**. A closeup of the 70% relative density W with single large pore at 15% strain showing the twin and dislocation formation in the [$\bar{1}$11] direction.

Fig. 9 shows the BDA analysis for defects in 70% relative density, multiple smaller nanopores under compression. Similarly to the single large nanopore system, there are very few defects present in the system for strains less than 10%. Small, unidentifiable defects begin forming at strains between 0% and 10% which are confined to the volume that connects the nearest pores in either the [010] and [001] directions. This is different to the large pore system that does not show a uniaxial defect confinement and occurs at lower strain values. Similarly to the large pore system, there are no defects in the volume connecting pores that are normal to the strain direction of [100]. When strain values of 10% are reached, twin boundaries and dislocations begin to form in the volume that is the intersection between the pores that are confined in the [010] and [001] directions. Pores also begin to deform inward in the strain direction [100]. Looking at the stress strain curve for this system, the loss in stress corresponds with the collapse of the perfect bcc structure surrounding the pores, however the collapse is confined to where the defects were confined in the previous strain values between pores in either [010] or [001] direction. The formation of the twins and dislocations follow similar formation mechanisms as the large single nanopore system where the formation of the twins and dislocations in the {211} slip planes in the <111> direction. With further increasing strain, the pores experience collapse and are observable in Fig. 9c where the pore surface in the [010] and [001] directions deform inward while the pore surface perpendicular to the strain direction remains in ellipsoid curvature. As strain continues to increase, the pore undergoes more severe deformation and the unaffected volume between pores parallel with the strain direction shrinks as the disordered volume grows and dislocations spread in the perfect bcc volume.

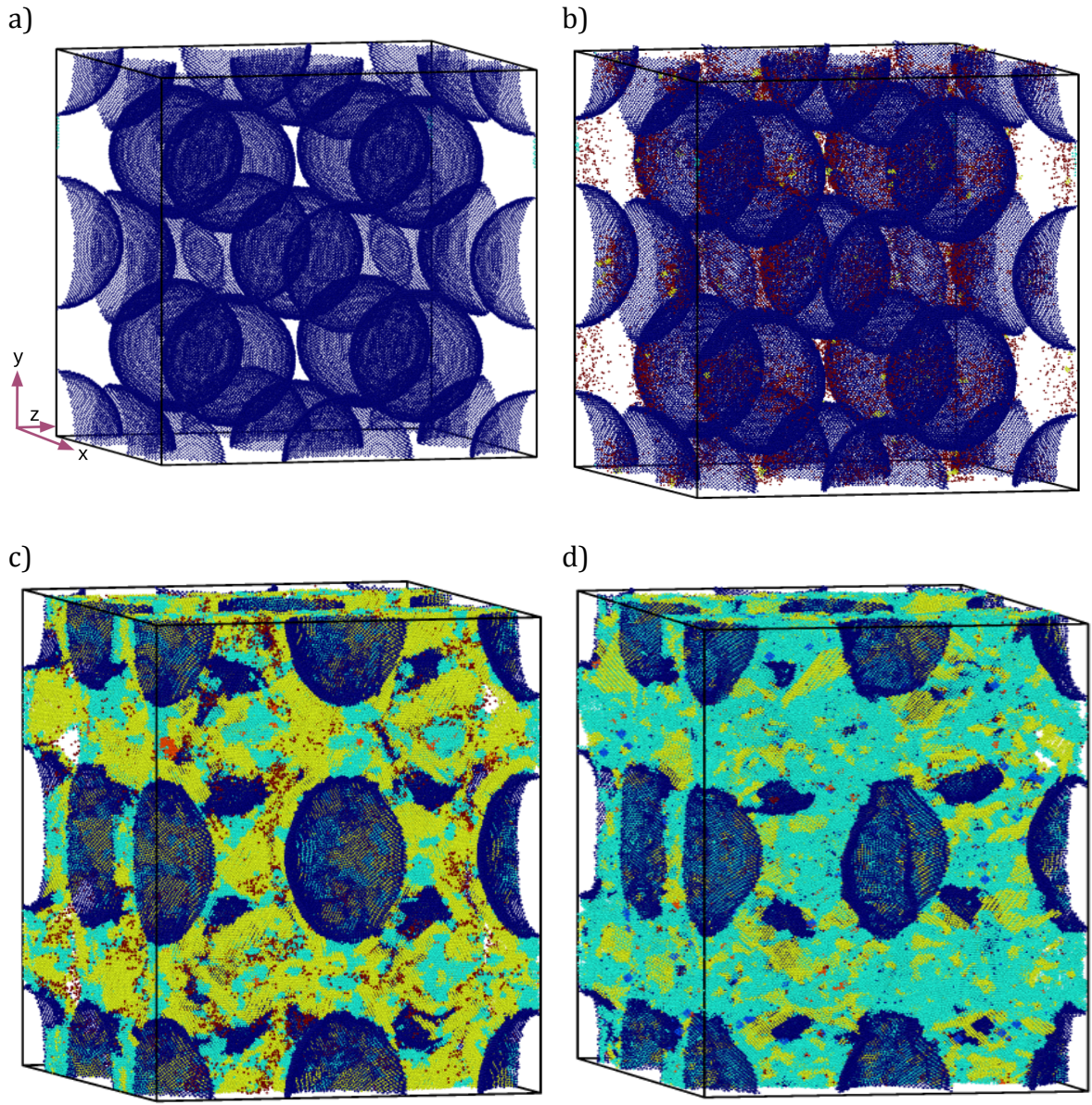

**Figure 9**. 30% relative density W with multiple small pores for strain values of (a) 0%, (b) 10%, (c) 15%, (d) 25%. Perfect bcc atoms are removed for visualization purposes.

At all strain values observed, there was no observed fracturing of the ligaments. This compares well with experiments on nanoporous metallic glass that observed ductile compression without any noticeable shear banding events [62]. The 46.5% relative density b-NP W system is analyzed after relaxation to ensure that no defects were in the initial system. This can be seen in Fig. 10a where the defect analysis only identifies surface atoms as the only existing defect in the microstructure. Similarly to the other nanopores structures, there are no additional defects produced until around a strain of 10% where defect structures begin to form in the larger ligament and nodes. The 10% strain is approximately the value where the yield stress is reached, showing plastic deformation through a rapid increase in defect concentration. Investigating the defect structures reveals a mix of 1/2<111> dislocations and <111>{112} twin boundaries forming together. The dislocations are shown to primarily nucleate at the surfaces and permeate through the bulk by staying near the surface. Deeper in the bulk, twin boundaries are seen to form by being connected by dislocations. These dislocations are 1/6<111> twinning dislocations that are adjacent on {112} planes that assist in the vertical growth of the twin boundaries deeper in the nanoporous structure [63]. These twins nucleating on 1/6<111> dislocations come from the dissociation of 1/2<111> screw dislocations into three twin dislocations as the energy favorable configuration under high stresses. Interacting twins can be observed intersecting in the [100] strain direction at 10% strain in Fig. 11a. An additional form of twin boundary intersection is formed at higher strains (Fig. 11b) where the twin boundaries are not in the form of a sharp 120% angle, but instead are shown to continuously curve, changing the twinning direction in the x-direction. These types of twin boundary formations are similar to unstable and inclined twin boundaries found in tungsten nanocrystals that play an important role in self-detwinning, leading to material deformation recovery [64]. A feature that was relatively absent from the pores system is the inclusion of vacancy clusters in the bulk of the b-NP W and is seen in Fig. 11b. The vacancy clusters congregate around the existing defect structures of dislocations and twin boundaries in the center of the ligaments at higher strain values. Vacancy clustering in the bulk around twin boundaries follows work conducted by Wang et al. that predict a decrease in vacancy formation energies around all grain boundary types up to 6 Angstrom away from the boundary region, allowing for easier formations of vacancies and vacancy clustering around the boundaries [65]. Planar faults appear at higher strain values of 20% strain or greater and must be joined together to a dislocation and either the surface or a twin boundary. This matches well with other bcc crystal structured nanopillar analysis. Yang et al. found that in a-Fe nanowires under compressive loading produces planar faults that stem from the accumulation of individual <111>/{112} twin boundaries and are one of the main driving forces for shape recovery during unloading [66]. The bicontinuous nanoporous systems show many different mechanisms demonstrating unloading recovery, leading to higher levels of shape recovery compared to the other systems analyzed that lack the particular twin boundary interactions.

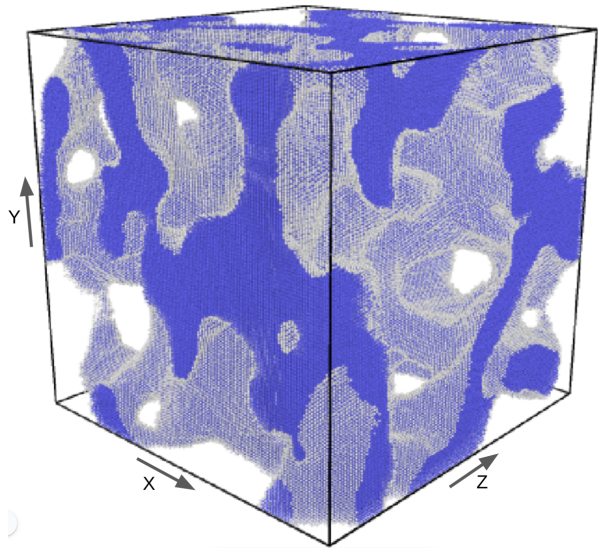
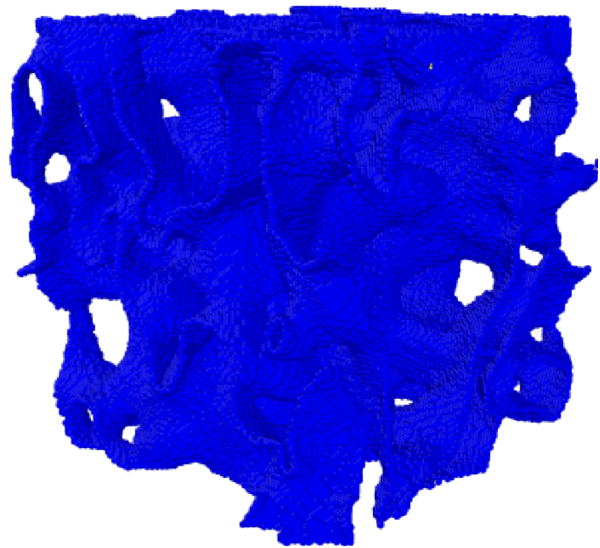

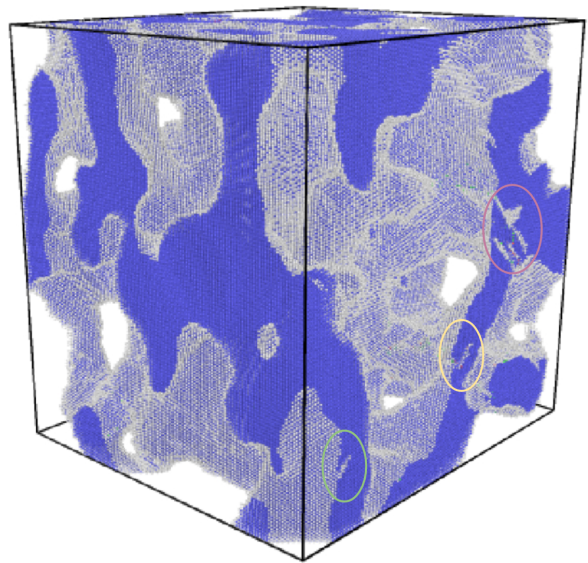
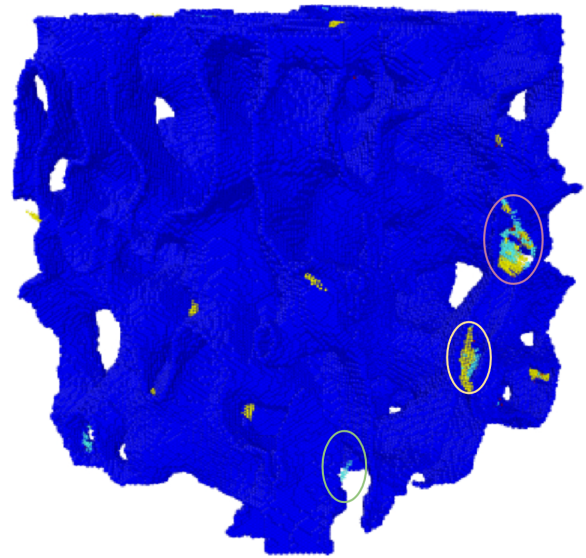

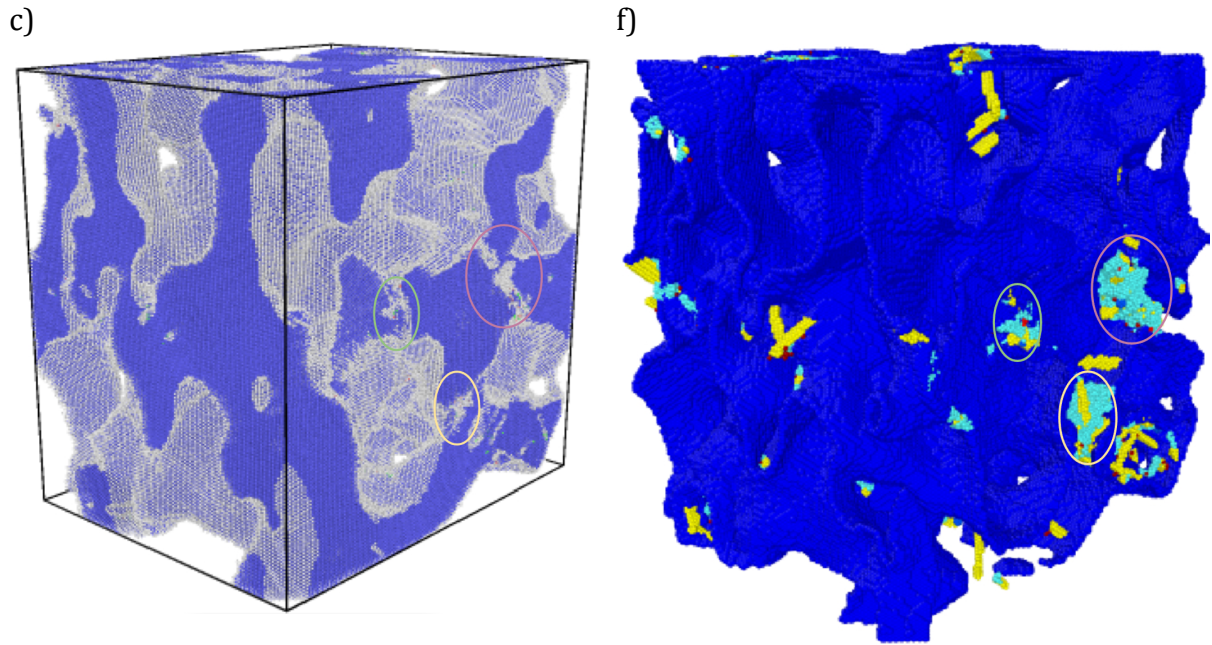

**Figure 10**. Compression of 46.5% relative density b-NP W with (a, d) 0% strain, (b, e) 10% strain, and (c, f) 20% strain. Left column figures are the nanoporous structures based off of nearest neighbor analysis that shows perfect bcc atoms as blue and non-bcc atoms as gray. Right column figures are analyzed by the BDA tool to analyze the internal structure with defects color coded as stated in the defect analysis. Colored circles between left and right columns outline the locations of the defects.

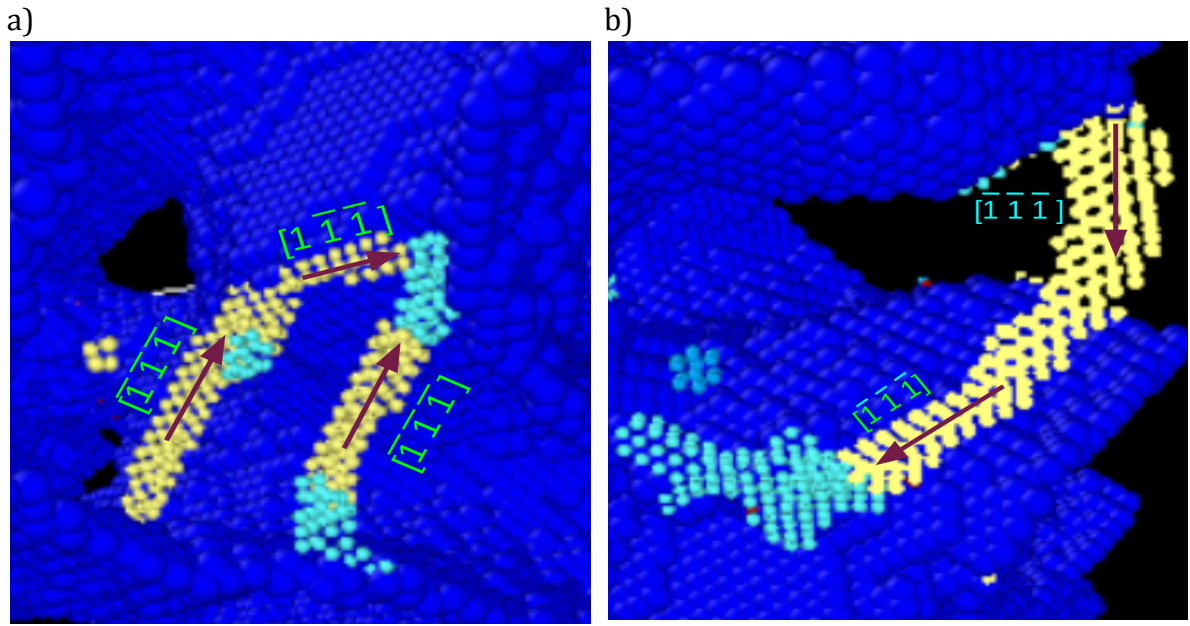

**Figure 11**. Enhanced portions of the 46.5% relative density NP-W at (a) 10% and (b) 20% strain displaying the fused and curved twin boundaries with dislocations and vacancy clusters.

IV. Conclusion

This paper used MD simulations to perform compression analysis on multiple W systems including bulk, NC, nanoporous, and bicontinuous nanoporous with emphasis on understanding the mechanical properties on the nanoporous systems. The system deformation is analyzed through system type and temperature ranging from room temperature 298 K to potential nuclear fusion reactor operating temperature 1150 K. Defect analysis was used to determine defect concentrations at specific strain and temperatures as well as the connection between defects and mechanical properties. This study determined that the b-NP W systems with low relative density have many attractive features for being a thermally stable material for use in fusion reactors, particularly very low defect concentrations compared to bulk and polycrystalline samples and unique twin boundary formation that facilitates shape recovery upon unloading. This translates to a high thermal stability of the b-NP W, allowing for easier expected operation under fusion reactor operating conditions.

**Acknowledgements**

This work was supported by the Nuclear Regulatory Agency [grant number 31310019M0045] that provided funding for Graduate Education.